\documentclass{article}

\usepackage{arxiv}

\usepackage{algorithmic}
\usepackage{graphicx}
\usepackage{textcomp}

\usepackage{enumitem}
\usepackage{wrapfig}
\usepackage{dirtytalk}
\usepackage{multirow}

\usepackage{algorithm}
\usepackage{multirow}
\usepackage{subcaption}
\usepackage{colortbl}

\usepackage[absolute]{textpos}
\usepackage{tcolorbox}

\usepackage{natbib}

\usepackage{wrapfig}

\definecolor{light-gray}{gray}{0.90}
\definecolor{dark-gray}{gray}{0.15}
\definecolor{lg-sep}{gray}{0.93}

\usepackage[font=small,labelfont=bf,tableposition=top]{caption}
\DeclareCaptionLabelFormat{andtable}{#1~#2  \&  \tablename~\thetable}

\usepackage{listings}
\makeatletter
\let\old@mkpream\@mkpream
\def\@mkpream{%
\ifx\CT@drsc@\relax\else\let\CT@drsc@ @\fi
\let\CT@arc@\relax
\old@mkpream}
\makeatother

\title{Applying Machine Learning on RSRP-based Features for False Base Station Detection}

\author{ {Prajwol Kumar Nakarmi} \\
    Business Area Networks, Ericsson\\	
    Stockholm, Sweden \\
	\texttt{prajwol.kumar.nakarmi@ericsson.com} \\
	\And
	{Jakob Sternby} \\
	Security Research, Ericsson\\	
    Lund, Sweden \\
	\texttt{jakob.sternby@ericsson.com} \\
  \And
	{Ikram Ullah}\thanks{Work conducted while at Ericsson} \\
	Business Area Networks, Ericsson\\	
    Stockholm, Sweden \\
	\texttt{ikram.lums@gmail.com} \\
}

\begin{document}

\maketitle

\begin{textblock}{12.5}(2,0.3)
  \begin{tcolorbox} [colback=white,colframe=black, sharp corners, boxsep=0mm, boxrule=0.2mm]
    \footnotesize The paper is to be published in the 17th International Conference on Availability, Reliability and Security (ARES 2022), August 23-26, 2022, Vienna, Austria. Please cite that version.     
    \\ Publication rights licensed to ACM. ACM ISBN 978-1-4503-9670-7/22/08. https://doi.org/10.1145/3538969.3543787.
  \end{tcolorbox}
\end{textblock}

\begin{abstract}    
    False base stations -- IMSI catchers, Stingrays -- are devices that impersonate legitimate base stations, as a part of malicious activities like unauthorized surveillance or communication sabotage. Detecting them on the network side using 3GPP standardized measurement reports is a promising technique. While applying predetermined detection rules works well when an attacker operates a false base station with an illegitimate Physical Cell Identifiers (PCI), the detection will produce false negatives when a more resourceful attacker operates the false base station with one of the legitimate PCIs obtained by scanning the neighborhood first. In this paper, we show how Machine Learning (ML) can be applied to alleviate such false negatives. We demonstrate our approach by conducting experiments in a simulation setup using the ns-3 LTE module. We propose three robust ML features (COL, DIST, XY) based on Reference Signal Received Power (RSRP) contained in measurement reports and cell locations. We evaluate four ML models (Regression Clustering, Anomaly Detection Forest, Autoencoder, and RCGAN) and show that several of them have a high precision in detection even when the false base station is using a legitimate PCI. In our experiments with a layout of 12 cells, where one cell acts as a moving false cell, between 75-95\% of the false positions are detected by the best model at a cost of 0.5\% false positives.
\end{abstract}


\section{Introduction}
A false base station \citep{3gpp33809} can be used as an attack tool that poses as a legitimate base station and performs various security and privacy attacks. As such, detection of false base stations is among the hottest topics in mobile networks security. In \citep{WhiteStingray}, the authors present their White-Stingray framework to assess the capabilities of different detection apps for the Android platform. Further, in \citep{Anatomy}, the authors have presented a survey on app-based, sensor-based, and network-based detection methodologies proposed by research studies and commercial products. 

One of the detection techniques proposed by researchers is to use radio frequency properties, with the premise that false base stations use cheap hardware with identifiable hardware imperfections. For instance, FBSleuth \citep{Zhuang2018FBSleuthFB} uses variance in the modulation errors, instantaneous frequency, and phases of the electromagnetic signals to generate fingerprints of false base stations.  A statistical approach has been devised by Ali and Fischer \citep{Ali2019EnablingFB}, who model the radio frequency noise as a univariate Gaussian distribution. The authors in \citep{Ali2019ThePN} used the idea of cellular network-based synchronization to look for higher carrier frequency offset values. Another detection technique involves using location information. e.g., Huang et al. \citep{Huang2018IdentifyingTF} propose using the location of mobile phones and legitimate base stations for checking the received signal strength, taking into consideration path loss, shadowing effect, and small-scale fading. 

In general, detection techniques that are not network-based are either not reliable due to lacking up to date network topology information; or not scalable due to additional hardware requirements. Hence, in this paper, we focus on the network-based approach that uses 3GPP standardized measurement reports and has been adopted by the 5G security standard \citep{3gpp33501}. Using this approach, the validity of Physical Cell Identifiers (PCIs) present in the measurement reports can be analyzed and is an effective rule-based detection technique \citep{MURAT}. This technique works well when the attacker, for stability of attack \citep{rupprecht19layertwo}, operates the false base station with a PCI that is not a legitimate neighbor. Nevertheless, a more advanced attacker might scan the neighborhood and determine all the  legitimate PCIs, and then operate the false base station with one of those legitimate PCIs. In such a case, the rule-based detection will produce false negatives. 

Therefore, we investigate Machine Learning (ML) techniques that can be applied to provide the necessary discriminatory power to fill the gap where the rule-based detection technique fails. Jian et al. in \citep{Jin2019RogueBS} describe using Gradient Boosting Machine algorithm on features like Reference Signal Received Power (RSRP) and information elements in System Information Block. Since this work uses supervised learning and treats false base station detection as a classification problem, it is unlikely to be practical because hardware and software configurations of both the false and legitimate network will vary widely in reality. Furthermore, regulatory issues often make it infeasible to obtain labeled malicious data needed for supervised learning. In \citep{Thuan2016StrengtheningMN}, the authors use anomaly detection algorithm named Seasonal Hybrid ESD (S-H-ESD) on features like handovers, location updates, and International Mobile Subscription Identity to International Mobile Subscription Identity mappings. Although this work addresses the issue as an anomaly detection problem, it requires more varied data sources, which limits its practicality. Further, to our knowledge, there is no work yet that aims at detecting a false base station using a legitimate PCI.

To that end, the aim of our work is to apply ML and make the network-based detection technique -- using 3GPP standardized measurement reports \citep{3gpp33501,MURAT} -- robust even when the false base station uses a legitimate PCI. The basic intuition behind our work is that a false base station affects the surrounding radio environment. It is so because one of the techniques that false base stations utilize is to provide a higher-than-normal signal strength to the mobile phones (e.g., by getting physically closer) in order to entice them to connect to it. For training and testing, we used the measurement reports from a simulated LTE network using ns-3 \citep{ns3home} which is a free and publicly available discrete-event network simulator (\S~\ref{sec:sim}). We did feature engineering based on the measurement reports and explored three different options, i.e., \emph{COL}, \emph{DST} and  \emph{XY} (\S~\ref{sec:features}). We evaluated four ML models, i.e., Regression Clustering, Anomaly Detection Forest, Autoencoder, and RCGAN (\S~\ref{sec:models}). The results we obtained indicate that applying ML is an effective technique that can be added to the arsenal of network-based detection of false base stations (\S~\ref{sec:res}). In our experiments with 12 cells where one of the cells acts as a false base station, the strongest ML model, ADF, detects 75-95\% of the false positions at a false positive ratio of $0.5\%$. 

Our main contributions are:
\begin{itemize}  
	\item We describe a ns-3 based LTE simulation setup for validating the effectiveness of false base station detection techniques.
	\item We present three different features derived from RSRPs in measurement reports and base station locations.	
	\item We provide and evaluate four ML models -- without the need for malicious data during training -- for detecting false base stations even when the attacker is using one of the legitimate PCIs. One of the models (based on Regression Clustering) is novel.
	\item We provide extensive experiments evaluating the performance of one ML model per serving cell and also include  aggregated results from multiple serving cells.
\end{itemize}

\section{Simulation and Data \label{sec:sim}}

\subsection{General}
The intuition behind our approach to using ML for detecting a false base station is that the presence of the false base station generates deviating data in the measurement reports received by the legitimate base station from mobile phones. Architecture wise, we build on the system components from \citep{MURAT}. Our primary data are the measurement reports on the network side that contain PCIs, Reference Signal Received Power (RSRP), and Reference Signal Received Quality (RSRQ) of neighboring cells. RSRP and RSRQ respectively represents strength and quality of received signals from a cell \citep{3gpp36214}. Our auxiliary data is the cell topology that consists of location of cells (cartesian coordinates in ns-3). We simulate the system using ns-3 as described below.

\subsection{ns-3 Simulation}
\begin{wrapfigure}{r}{0.35\linewidth}
    \centering   
    \vspace{-5pt}
    \includegraphics[width=0.35\textwidth]{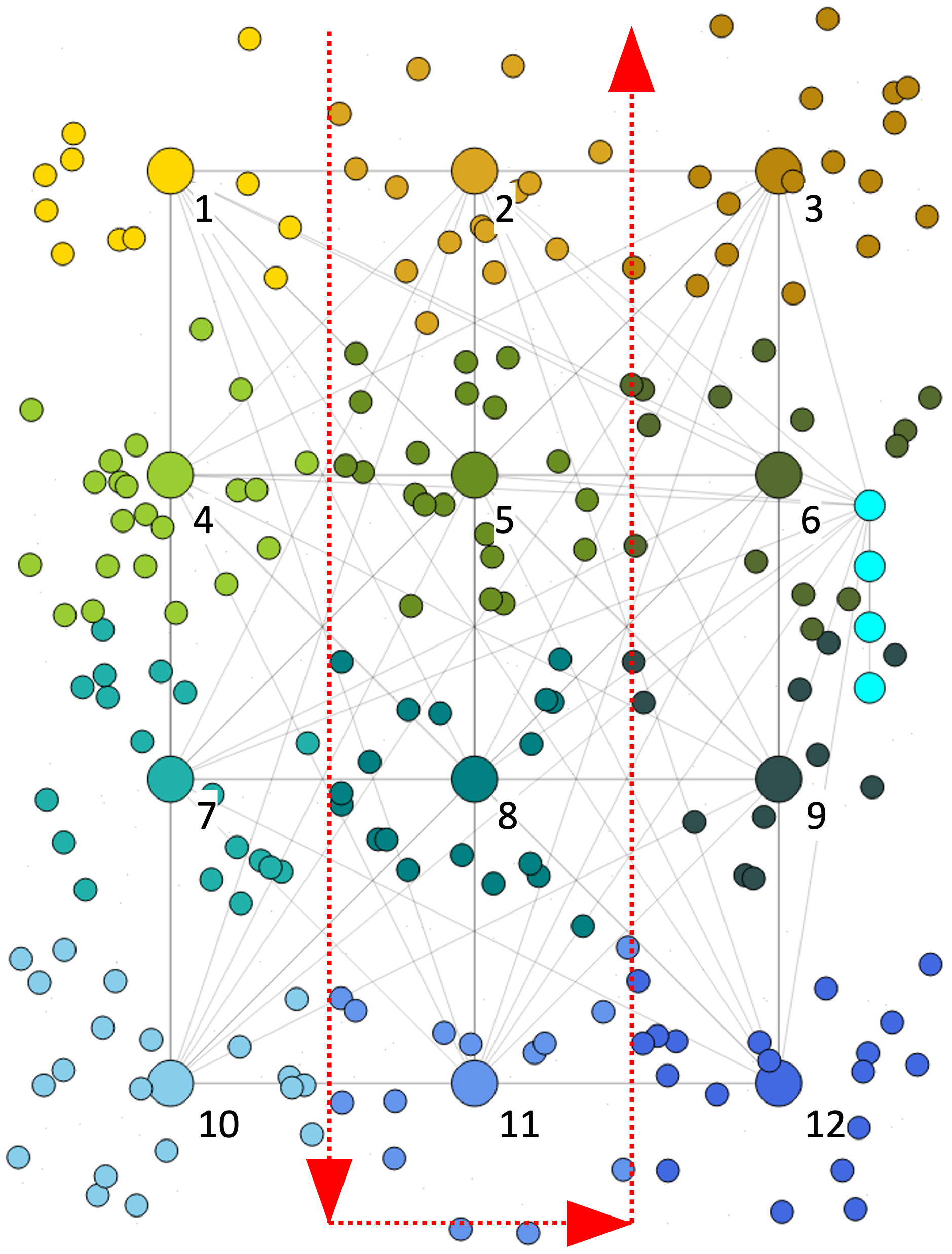}
   \caption{ns-3 Simulation Layout.\label{fig:layout_city}}
\end{wrapfigure}

Using the LTE module in ns-3, we simulated a network with 12 cells in a 3 x 4 layout as shown in Figure \ref{fig:layout_city} (bigger circles represent cells; smaller circles represent mobile phones; and Cyan colored circles represent core network functions). ns-3 uses the Cartesian coordinate system. 

We setup the cells layout using the Grid Position Allocator with DeltaX=500, DeltaY=500, GridWidth=3, and LayoutType=RowFirst. We set the propagation loss model to Okumura Hata \citep{ns3oh}, number of RA preambles to 10, and SRS periodicity to 320. The mobile phones moved within the specified boundary (-250, 1250), (-250, 1750) according to the RandomWalk2dMobilityModel in which they moved with a specified speed of 5 m/sec and direction chosen at random every 20 seconds. We kept many other configurations in the ns-3's LTE module to their defaults, e.g., A2-A4-RSRQ is the handover algorithm which takes into account the RSRQ of serving cell getting worse (event A2) and that of neighboring cell getting better (event A4); transmission power of cell is 30 dBm. 

During generation of \emph{training} data, we created 200 mobile phones and kept all 12 cells operational with Automatic Neighbor Relation (ANR). All cells were placed at the height of 25 m and the mobile phones were placed at the height randomly chosen between 1.5 to 2 m. The simulation was run for 1000 seconds with random seed set to 1. 

For \emph{test} data generation, ideally, we would have to introduce a new cell with existing PCI. But this was not supported by ns-3 because ns-3 natively avoids PCI collision by automatically assigning PCIs up to 65,535 distinct PCIs. Therefore, we envision a story as follows. We assume that one of the cells has been decommissioned, but the cell topology is not yet updated. The attacker capitalizes on this and starts its own false cell with the PCI of the decommissioned cell and moves along a path. Mind that even though we could not simulate the exact scenario we wanted with PCI collision, the final effect is that the radio conditions (RSRP and RSRQ) will be affected and our main goal still remains valid in the sense that we want to detect changes in the radio conditions even though there is no new PCI. 

We generated separate testing data with each cell as decommissioned (which we achieved by setting AdmitHandoverRequest to false). The attacker's cell started at the original position of the decommissioned cell; stayed there for 200 seconds (but at a lower height of 2 m assuming that the attacker had its equipments in a van); and after that, moved for 120 seconds along the red lines shown in Figure \ref{fig:layout_city}, i.e., (250, -250), (250, 1750), (750, 1750), and (750, -250). We note that the idea behind the movement of the false cell in the simulation is to contribute to the difference in radio environments and is irrelevant otherwise to our detection technique. 

We created 100 mobile phones and ran the test simulation with random seed 9. We generated \emph{validation} data in the same way as the \emph{testing} data except for using random seed 1 and only using the 200 seconds (without the cells moving).

\subsection{Training and Testing Data\label{sec:train_test_data}}
We preprocessed the data by removing records without any neighbors as these cannot contain false base station information. We separated the rest of the data according to the specified serving cell of each record. This way, for each serving cell, we obtained one training set, and 11 test and validation sets for each simulation where one of the other cells in the layout was moving.

\begin{figure}[]
    \centering       
    \raisebox{-\height}{\includegraphics[width=0.3\textwidth]{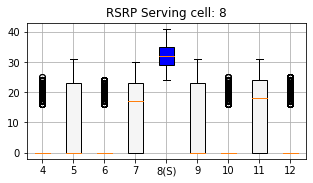}} 
    \raisebox{-\height}{\includegraphics[width=0.3\textwidth]{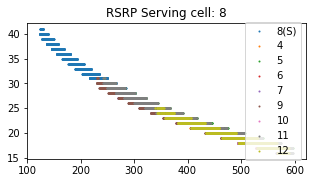}}
    \raisebox{-\height}{\includegraphics[width=0.3\textwidth]{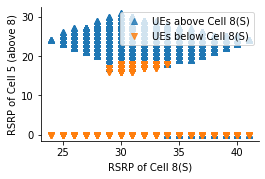}}
    \raisebox{-\height}{\includegraphics[width=0.3\textwidth]{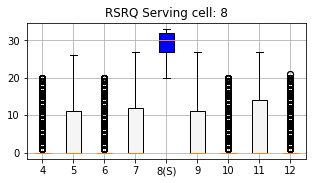}}
    \raisebox{-\height}{\includegraphics[width=0.3\textwidth]{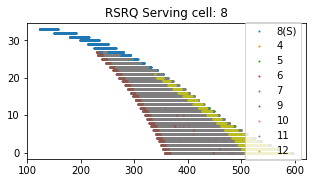}}
    \raisebox{-\height}{\includegraphics[width=0.3\textwidth]{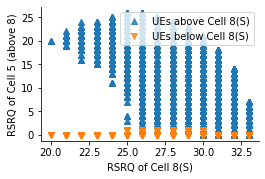}}       
    \caption{Training Data Dynamics\label{fig:train_data}}   
\end{figure}

Figure~\ref{fig:train_data} shows the dynamics of the training data generated by our ns-3 simulation. For clarity of plots, we only chose the serving cell in the middle, i.e., cell 8. The first column shows two box plots RSRPs and RSRQs of all the cells reported to cell 8 (marked with S in the label and blue color in the plot). In general, it can be seen that the serving cell (8) has higher RSRP/RSRQ followed by the immediate neighboring cells (5, 7, 9, 11). It is also observable that although the serving cell clearly has better RSRQ, the RSRQs of neighboring cells are more erratic compared to the RSRPs. The second column depicts how the reported RSRPs/RSRQs vary with the distance of mobile phones with each cells. There is a visibly strong negative correlation, which is also intuitive, that the power and quality decrease with increasing distances. However, the RSRQ plot is more erratic. The third column shows plots of RSRP and RSRQ correlation between the serving cell (cell 8) and a cell above it (i.e., cell 5). Strong correlations are not clearly visible, except that, many a times, for the UEs below the cell 8 (orange color), the RSRP/RSRQ of cell 5 is 0. This may be attributed to how Okumura Hata propagation loss model works.

\begin{figure}[]
    \centering       
    \raisebox{-\height}{\includegraphics[width=0.3\textwidth]{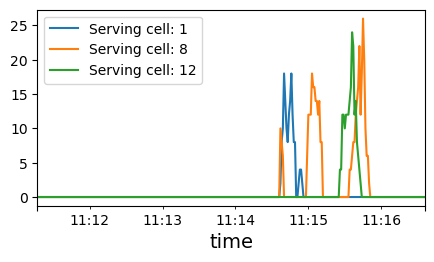}}        
    \raisebox{-\height}{\includegraphics[width=0.3\textwidth]{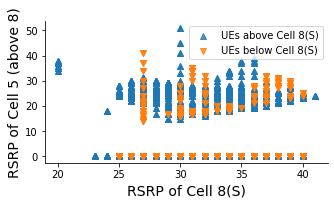}}
    \raisebox{-\height}{\includegraphics[width=0.3\textwidth]{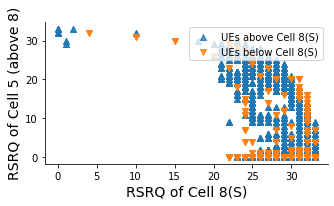}}    
    \caption{Testing Data Dynamics\label{fig:test_data}}   
\end{figure}

Some dynamics of the testing data are depicted in Figure~\ref{fig:test_data}. The first figure shows how many measurement reports become anomalous from the perspective of serving cells 1, 8, and 12, when cell 5 moves. The other two figures show the plots of RSRP and RSRQ correlation between the serving cell 8 and cell 5 above it. It is clearly visible that the dynamics are not as they used to be for the training data (Figure~\ref{fig:train_data}).

\section{Feature Engineering}\label{sec:features}
We extract robust features suitable for applying ML to detect false base stations. From the measurement reports, the RSRQ values are excluded because they are erratic, as shown in Figure~\ref{fig:train_data}. We combine the RSRP values of cells with the locations of those cells to build a model of how these values typically vary around each serving cell. When a false base station, disguised with a legitimate PCI, appears in a measurement report, the RSRP values reported for that PCI will typically be different than if the PCI belonged to the legitimate base station with the same PCI. 

Since the serving cells are stationary, we use the data collected by each cell to create a model of the reported neighbor values. We have explored three different options for extracting data to train models of neighbor RSRP values around each serving cell. Each option includes the RSRP of the serving cell, the number of neighbors, and the RSRP of the existing neighbors in one of the following ways:

\begin{enumerate}
\item[\textbf{COL}] One fixed column per neighbor PCI with the RSRP-value of that neighbor.
\item[\textbf{DST}] Two columns of mixed neighbor values per existing neighbor, one for the distance of the neighbor from the serving cell and one for the RSRP of the neighbor. The maximum number of DST features therefore be twice the number of concurrent neighbors in any report.
\item[\textbf{XY}] Same as \emph{DST} but containing two columns modelling the cartesian coordinates of the neighbor relative to the serving cell instead of just the distance. The number of neighbor derived features will here be three times the maximum number of concurrent neighbors in any report.
\end{enumerate}  

Note that all of these features will contain missing elements as the number of neighbors in reports varies. The feature sets \emph{DST} and \emph{XY} enable mixing different neighbors in the same column by adding additional positional features to distinguish different neighbor PCIs instead of extracting the RSRP value of each neighbor PCI into separate features. The \emph{COL} feature does not mix values from different neighboring cells and thereby has the greatest capability in modelling exact correlations between the RSRP values of the serving cell and specific neighbors. However, in a setting where the maximum number of concurrent cells is much smaller than the number of different neighbors this may not work as well due to the quickly growing set of missing values in the data matrix as seen in Figure~\ref{fig:sorted_feat}.

\begin{figure}[H]
	\begin{subtable}[h]{0.2\columnwidth}
	\centering
		\begin{tabular}{c c c c}
	\hline
	\hline
	\rowcolor{dark-gray}
	\textcolor{white}	2 & \textcolor{white}3 & \textcolor{white}4 & \textcolor{white}5\\
	\hline
	\hline
		45 & -  & 47 & - \\

	\doublerulesepcolor{lg-sep}
	\rowcolor{light-gray}
		 -  & -  & 46 & - \\
	\doublerulesepcolor{nearwhite}
		47 & -  & 47 & -\\
	\rowcolor{light-gray}
	\doublerulesepcolor{lg-sep}
		 -  &  - &  - & 40 \\
	\doublerulesepcolor{nearwhite}
		 -  & 40 & - & -\\
	\hline
	\end{tabular}
	\label{tab:feat_col}
	\caption{\emph{COL}}
	\end{subtable}
	\hspace{0.5cm}
	\begin{subtable}[h]{0.3\columnwidth}
	\centering
	\begin{tabular}{c c c c}
	\hline
	\hline
	\rowcolor{dark-gray}
		\textcolor{white}{RSRP} & \textcolor{white}{DST} & \textcolor{white}{RSRP} & \textcolor{white}{DST}\\
	\hline
	\hline
		47 & 1000 & 45 & 1000 \\
	\rowcolor{light-gray}
		46 & 1000 & - & -\\
		47 & 1000 & 47 & 1000 \\
	\rowcolor{light-gray}
		40 & 1414 & -  & - \\
		40 & 2000 & - & - \\
	\hline
	\end{tabular}
	\label{tab:feat_dst}
	\caption{\emph{DST}}
	\end{subtable}
	\hfill
	\begin{subtable}[h]{0.5\columnwidth}
	\centering
	\begin{tabular}{c c c c c c}
	\hline
	\hline
	\rowcolor{dark-gray}
		\textcolor{white}{RSRP} & \textcolor{white}{X} & \textcolor{white}{Y} & \textcolor{white}{RSRP} & \textcolor{white}{X} & \textcolor{white}{Y} \\
	\hline
	\hline
		47 & 1000 & 0 & 45 & 0 & 1000 \\
	\rowcolor{light-gray}
		46 & 0 & 1000 & - & - & - \\
		47 & 1000 & 0 & 47 & 0 & 1000 \\
	\rowcolor{light-gray}
		40 & 1000 & 1000  & - & - & -\\
		40 & 2000 & 0 & - & - & -\\
	\hline
	\end{tabular}
	\label{tab:feat_xy}
	\caption{\emph{XY}}
	\end{subtable}

\caption{The same records shown with the three different examined methods for feature extraction, i.e., \emph{COL}, \emph{DST} and \emph{XY}.\label{fig:sorted_feat}}
\end{figure}

\section{ML Models for Novelty Detection \label{sec:models}}

Novelty detection usually implies semi-supervised learning, where data from the normal class (i.e., one known label) is used to train a model. This is then used as a reference when determining if new input belongs to the training distribution or if it is previously unseen (e.g., a novelty). In our experiments we have evaluated one new model based on Regression Clustering and three existing novelty detection methods based on Anomaly Detection Forest (ADF) \citep{jakob2020adf}, traditional Autoencoders (AE) \citep{autoencoder2020}, and Regularized Cycle Consistent Generative Adversarial Network (RCGAN) \citep{yang2020regularized}. 

\textbf{Regression Clustering}. In Regression Clustering \citep{regresscluster}, a data set is first grouped into different clusters and regression models are trained for each cluster in order to be able to predict values. We use the \emph{COL} features, and build clusters and regressor models on the training data, such that RSRP of a cell (y) is trained using the RSRPs of other cells around that cell (x). Then, during the test phase, the trained models are used to predict the RSRP of the cell. If the predicted RSRP differs significantly from the observed RSRP, we determine that there is some anomaly. Now, when does such anomaly occur (i.e., when is the predicted RSRP value is not close to the observed RSRP value)? It is when a false base station pollutes the data. In other words, if there is a false base station, the prediction will not be correct; and if a false base station is absent, the prediction will be correct. We can use this intuition and iterate through cells. Then, a cell in whose absence the prediction becomes correct is the false cell. Training logic is shown in Algorithm~\ref{alg:regress_train} and test logic is shown in Algorithm \ref{alg:regress_test}. We used KMeans for clustering and Random Forest for regression from scikit-learn \citep{sklearn}.

\textbf{ADF}. ADF is an ensemble of binary decision trees built from subsets of the training set selected at random. The training consists of two phases where the first phase splits a training subsample into smaller regions in input space and the second phase sets feature-wise anomaly borders for each region. The implementation of ADF is adopted from \citep{jakob2020adf} with hyperparameters: subsample size $|X| = 512$, anomaly margin $a = 1$, isolation level $\eta = 0.05$ and max depth $D = 14$. Each model contains 150 anomaly detection trees.

\textbf{AE}. Autoencoders is a common choice of algorithm for anomaly detection.
An autoencoder is defined by a self-mapping network that is trained by minimizing the reconstruction error on a training set. The basic motivation for using reconstruction error for anomaly detection is that encoding and decoding has been trained with a certain training distribution and thus the reconstruction may not work as well for elements that are dissimilar to the training distribution. The autoencoder architecture tested in our experiments used the same hourglass shaped structure presented in \citep{autoencoder2020} with three internal dense layers. The number of nodes in the hidden layers were $[0.8F, 0.5F, 0.8F]$, where $F$ is the number of input features (this varies with serving cell as well as the type of features used). Training of the autoencoder models was conducted with the Tensorflow framework using the Adam optimizer. The loss function was mean square error (MSE) and the batch size was 50. Each model was trained for 150 epochs with a starting learning rate of 0.002 updated every 50 epochs with a decay of 0.8.

\textbf{RCGAN}. RCGAN is one of several recent GAN-inspired methods for anomaly detection. The main components of the RCGAN is an Encoder $E()$ which is analagous to the first layers of the autoencoder and a Generator $G()$ which maps the encoded input back to input space (from encoded/latent space) and thereby corresponds to the final layers of the autoencoder. So the actual structure is very similar to the autoencoder but with some significant differences in the training process. The implementation of RCGAN in this paper used the bi-direcitonal GAN setup from ALAD \citep{zenati2018adversarially} but with an added latent penalty distribution as proposed in \citep{yang2020regularized}. The Encoder used one hidden layer with $0.7F$ number of nodes where $F$ is the number of input features and the latent dimension (the dimension of output from the Encoder) was $0.3F$. The Generator used two hidden layers with $[0.5F, 0.7F]$ nodes. The three discriminators $D_{xx}, D_{xz}, D_{zz}$ had one hidden layer each with $0.5F$ nodes. Training was conducted as in \citep{yang2020regularized} apart from not using exponential moving average or spectral normalization. Similar to AE, the Tensorflow framework was used to train the RCGAN models for 150 epochs with a starting learning rate of 0.002 updated every 50 epochs with a decay of 0.8.

\begin{minipage}{0.42\textwidth}
  \footnotesize  
  \begin{algorithm}[H]
    \caption{Regression Clustering: Training}\label{alg:regress_train}
    \begin{algorithmic}            
      \STATE $CELLS\gets range(cells)$
      \FOR{$C$ in $CELLS$} 
        \STATE $notC\gets CELLS - C$
        \FOR{$nC$ in $notC$}
          \STATE $x\gets notC - nC$
          \STATE $y\gets C$
          \STATE $CLUSTERS\gets$ cluster on $y$
          \FOR{$CL$ in $CLUSTERS$}
            \STATE Train $y_{CL}$ with $x_{CL}$
          \ENDFOR              
        \ENDFOR            
      \ENDFOR       
    \end{algorithmic}
  \end{algorithm}
\end{minipage}%
\begin{minipage}{0.58\textwidth}
  \footnotesize
  \begin{algorithm}[H]
    \caption{Regression Clustering: Test}\label{alg:regress_test}
    \begin{algorithmic}      
      \STATE $CELLS\gets range(cells)$
      \STATE $FLAG_{PASS\_1}\gets empty$
      \STATE $FLAG_{PASS\_2}\gets empty$
  
      \STATE /*Predict and calculate residues*/
      \FOR{$C$ in $CELLS$} 
        \STATE $notC\gets CELLS - C$
        \FOR{$nC$ in $notC$}
          \STATE $x\gets notC - nC$
          \STATE $y\gets C$
          \STATE $CLUSTERS\gets$ cluster on $y$
          \FOR{$CL$ in $CLUSTERS$}            
            \STATE Predict $yP_{CL}$ with $x_{CL}$
            \STATE $residueY\gets |y_{CL} - yP_{CL}|$
          \ENDFOR                    
          \STATE $FLAG_{PASS\_1}[C, nC]\gets residueY > THRESHOLD$
        \ENDFOR                  
      \ENDFOR 
      
      \STATE /*From perspective of each cell, flag others*/
      \FOR{$C$ in $CELLS$} 
        \STATE $FLAG_C\gets FLAG_{PASS\_1}[C, *]$
        \STATE $FLAG_{PASS\_2}\gets True$ if only one 0 in $FLAG_C$, $False$ otherwise
      \ENDFOR 
      
      \STATE $FLAG_{PASS\_2}$ contains the flagged/unflagged verdict
    \end{algorithmic}
  \end{algorithm}
\end{minipage}

\section{Results and Discussion \label{sec:res}}
Training and test data are described in Section~\ref{sec:train_test_data}. The number of records for each serving cell are listed in Table~\ref{tab:datasets}. 

To evaluate the success of the novelty detection methods we tuned the threshold on the anomaly scores for each method on the benign samples of the validation set such that only $0.5\%$ their scores were above the threshold. In Table~\ref{tab:results}, we show the recall of the true positives (ratio) with this threshold. 

\begin{figure}
	\centering
	\includegraphics[width=0.35\columnwidth]{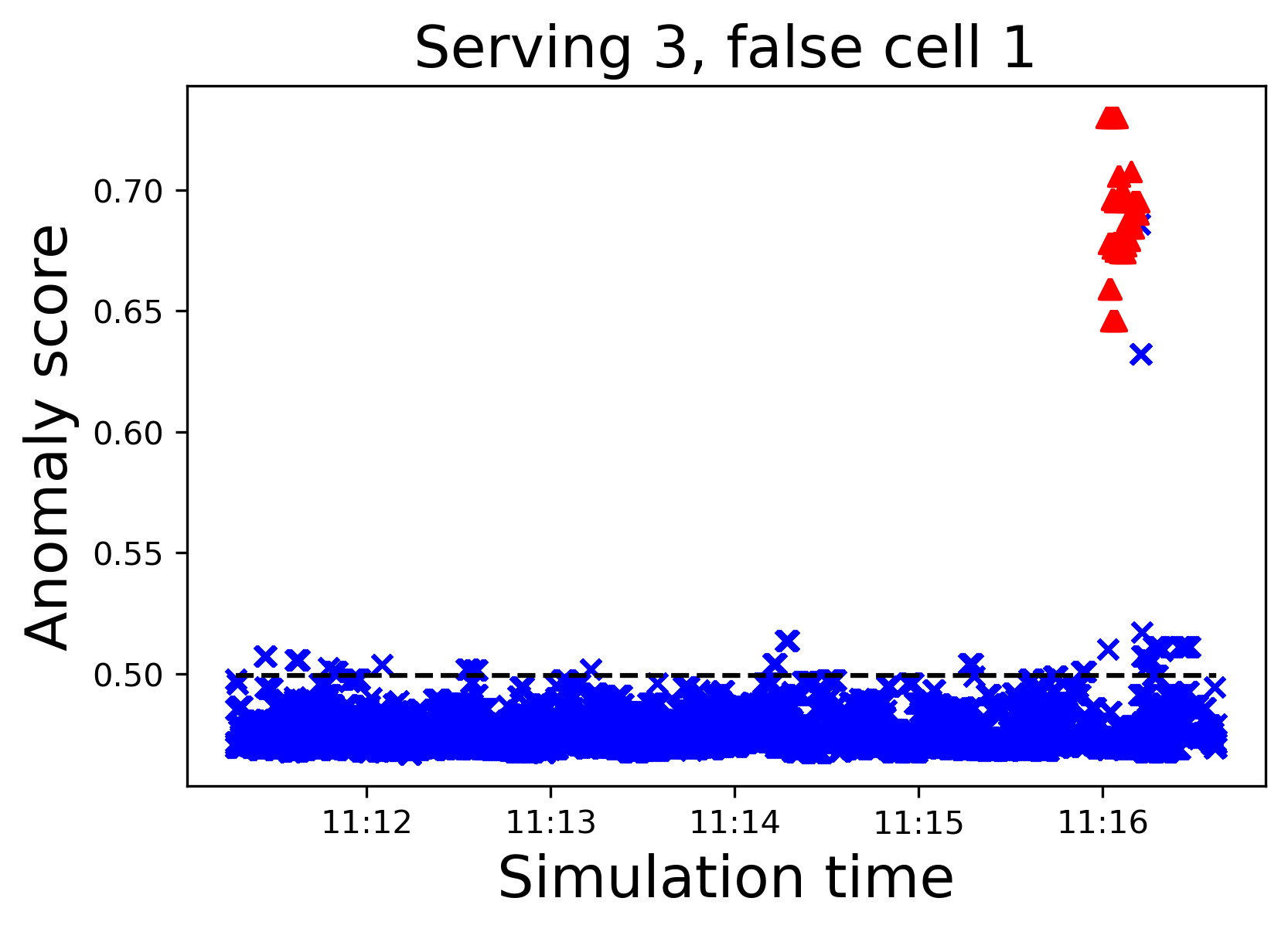}
	\includegraphics[width=0.35\columnwidth]{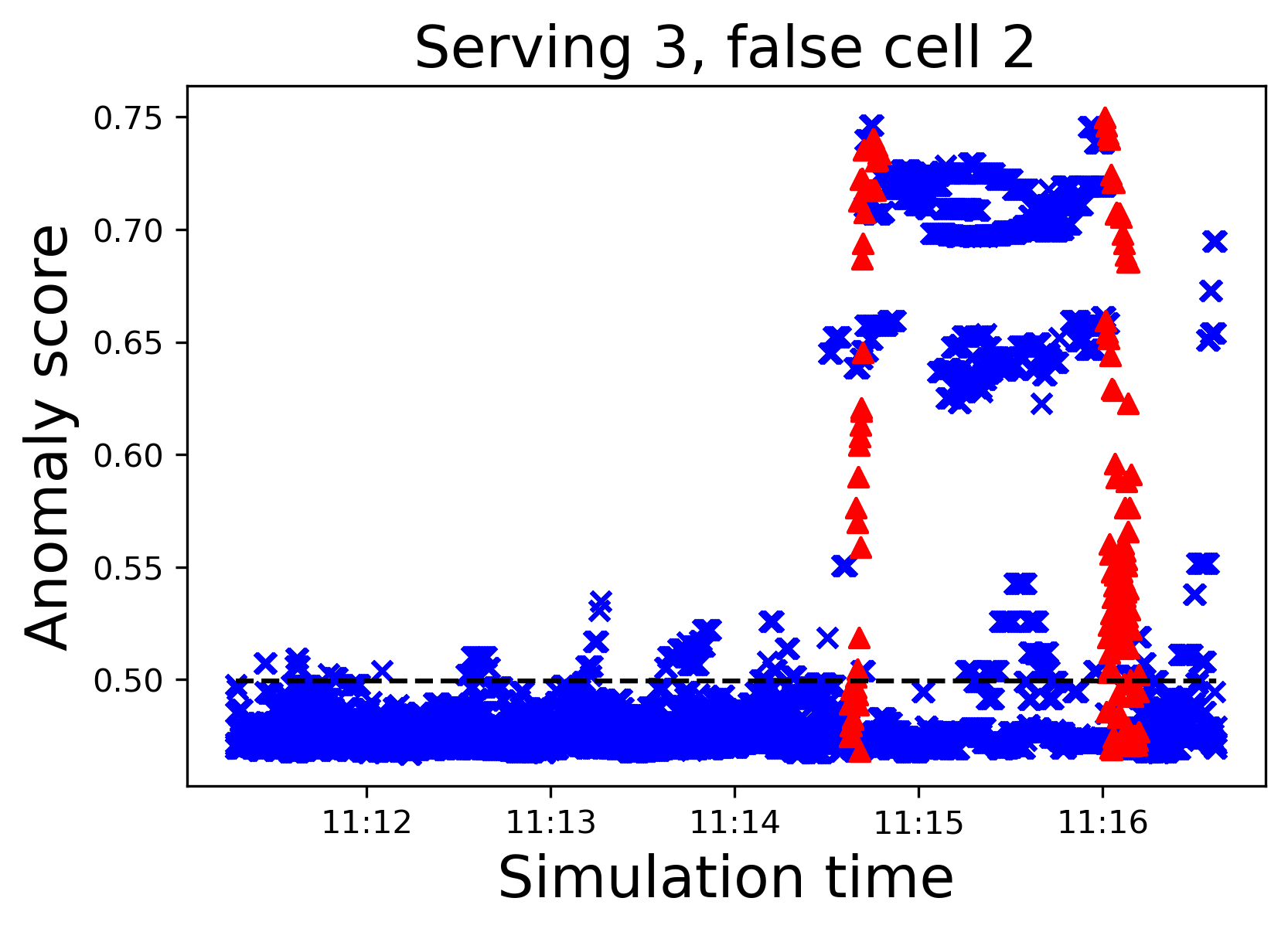}
	\caption{Anomaly scores for serving cell 3 for false cells 1 (left) and 2 (right). Red triangles are derived from measurement reports containing the moving cell.}
	\label{fig:serving3_scores}
	\end{figure}

Note that, as the false cell starts moving in the simulation, it leaves a vacancy such that surrounding serving cells start receiving measurement reports from mobile phones previously reporting to the moved cell and this causes novelties. This phenomenon is due to the limitation in ns-3 that PCI collision was not supported. The effect is clearly visible in Figure~\ref{fig:serving3_scores} that serving cell 3 gets false positives when cell 2 starts moving, but not when cell 1 moves since movement of cell 1 does not cause mobile phones previously reporting to cell 1 to report to cell 3. 

Furthermore, Figure~\ref{fig:serving3_scores} indicates that the presence of a false base station may be further corroborated by multiple mobile phones in different locations picking up the false cell and causing a cluster of measurement reports with higher anomaly scores. On the other hand, the false positives in the absence of a false base station seem to occur more isolated as would be the case for a limited number of phones showing up in a new location.

\begin{table*}[]
\footnotesize
\caption{Training, validation and test data sets}
\label{tab:datasets}
\centering
\resizebox{0.8\textwidth}{!}{%
\begin{tabular}{|c|c|c|c|c|cl|}
\hline
\hline
Serving cell & \multicolumn{1}{|c|}{Training size} & Neighbors in training & \multicolumn{1}{|c|}{Validation size} & \multicolumn{1}{|c|}{Test size} & \multicolumn{1}{|c}{Anomalies} & (Static) \\
\hline
\hline
	1 & 68153 & 2-5,7 & 77803 & 100799 & 3018 & (1156)\\
	2 & 52691 & 1,3-6 & 69101 & 116745 & 4886 & (1914)\\
	3 & 41913 & 1,2,5,6,9 & 81181 & 107461 & 2290 & (980)\\
	4 & 49203 & 1,2,5,7,8,10 & 40297 & 111351 & 2814 & (724)\\
	5 & 61545 & 1-9 & 38469 & 135363 & 4136 & (1112) \\
	6 & 48927 & 2,3,5,8,9,12 & 71975 & 103325 & 1220 & (196)\\
	7 & 41867 & 1,4,5,8,10,11 & 70967 & 54243 & 1204 & (272) \\
	8 & 53851 & 4-7,9-12 & 61001 & 65939 & 3106 & (642)\\
	9 & 65329 & 3,5,6,8,11,12 & 77111 & 86549 & 1570 & (434)\\
	10 & 36929 & 4,7,8,11,12 & 52769 & 90793 & 950 & (300)\\
	11 & 50461 & 7-10,12 & 45015 & 92313 & 3432 & (1514)\\ 
	12 & 59347 & 6,8-11 & 86957 & 89701 & 1682 & (696) \\
\hline
\hline
\end{tabular}%
}
\end{table*}

\begin{table*}[]
	\footnotesize
	\caption{Recall of TP for (Model(feat)) at benign FPR $0.5\%$ (FPR without static anomalies)}
	\label{tab:results}
	\centering
	  \resizebox{.96\textwidth}{!}{%
	\begin{tabular}{|c|c|c|c|c|c|c|c|}
	\hline
	\hline
	Serving cell & RCGAN(COL) & RCGAN(XY) & AE(COL) & AE(XY) & REGRESSOR (COL) & ADF(DST) & ADF(COL) \\
	\hline
	\hline
		1 & 65(43)\% & 78(65)\% & 52(22)\% & 71(53)\% &  49(17) \% & 79(65)\% & {\bf 81(69)\%} \\
		2 & 62(37)\% & 82(70)\% & 60(34)\% & 50(18)\% &  48(15)\% & 83(72)\% & {\bf 89(83)\%} \\
		3 & 74(55)\% & 89(81)\% & 50(13)\% & 76(59)\% &  49(11)\% & {\bf 91(84)\%} & 90(83)\% \\
		4 & 65(52)\% & 71(60)\% & 46(27)\% & 57(42)\% & 40(19)\% & 77(69)\% & {\bf 82(76)\%} \\
		5 & 57(42)\% & 79(71)\% & 50(32)\% & 50(31)\% & 34(9)\% & 86(80)\% & {\bf 86(81)\%} \\
		6 & 67(61)\% & 69(63)\% & 48(38)\% & 50(41)\% & 42(31)\% & 80(77)\% & {\bf 85(82)\%} \\
		7 & 63(52)\% & {\bf 80(74)\%} & 51(37)\% & 46(30)\% & 44(27)\% & 46(30)\% & 59(47)\% \\
		8 & 74(67)\% & 74(67)\% & 44(29)\% & 42(27)\% & 31(13)\% & 84(80)\% & {\bf 86(82)\%} \\
		9 & 77(67)\% & 85(80)\% &  48(29)\% & 52(34)\% &  44(23)\% & 87(82)\% & {\bf 92(88)\%} \\
		10 & 75(63)\% & 72(59)\% & 60(42)\% & 64(47)\% &  55(34)\% & {\bf 76(66)\%} & 70(56)\% \\
		11 & 76(58)\% & 75(55)\% & 61(30)\% & 54(18)\% &  54(18)\% & 80(63)\% & {\bf 87(77)\%} \\
		12 & 78(62)\% & 85(75)\% & 51(16)\% & 68(46)\% &  50(15)\% & 85(75)\% & {\bf 90(83)\%} \\
	\hline
	\hline
	\end{tabular}
	}
	\end{table*} 
	
\subsection{Data-derived Static Novelties}\label{sec:rulebased_anomalies}
During training of the ML models we obtain an updated list of present neighbors for each serving cell. Neighboring cells never seen for a serving cell during training will imply that these should be marked as suspicious if they start appearing during test (see Table~\ref{tab:datasets}). This way, we obtain data-derived static novelties to combine with our ML methods for detection. As these novelties are easy to detect, we note that the attacker would need to avoid inducing such novelties to evade detection.

\subsection{Serving Cell Results}\label{sec:results_serving}
One novelty detection model for each evaluated algorithm was trained on the training data without anomalies listed in Table~\ref{tab:datasets}. In order to evaluate the possibility of detecting false base stations using models built on these features in a realistic setting, we determined a threshold for the anomaly score based on an acceptable rate of false positives on a validation set of benign records. Although we evaluated more combinations of features and ML-algorithms, only a selection including the best performing combinations is listed in Table~\ref{tab:results} to avoid clutter. For these algorithms we show the novelty recall results when setting a threshold implying a $0.5\%$ false positive rate on the benign validation set. For each model of serving cells we show recall results both with static anomalies as well as with those excluded (in parenthesis) in Table~\ref{tab:results}.

The \emph{COL} features performed well in our simulated setting where the distinctly occurring neighbor cells is fairly few (we had 12 cells in total), but we believe that the \emph{XY} features could prove more useful in the diversity of real deployment settings (where there are hundreds of cells). Although the \emph{XY} features imply an extra attribute for every occurring neighbor, the added possibility to distinguish the direction of the neighbor seems to be superior to just using the distance of the \emph{DST} features. The RCGAN consistently performs better than the AE and we stipulate that this is due to the lack of guarantee that the AE will have high reconstruction errors for the anomalies as previously noted in \citep{zong2018deep}. The success of ADF may be owed to its even more explicit modelling of the benign distribution. Unlike AE and RCGAN which both perform better with the \emph{XY} features it does not contain any dimensionality reduction (encoding) which may be the explanation as to why the \emph{COL} features provides the best results for ADF. The Regressor does not perform well in this setting.

\subsection{Aggregated Serving Cell Results}\label{sec:agg_results}
In terms of being able to detect a false base station, the numbers for recall of true positives (the records where a moved/altered cell occurs as a neighbor) in Table~\ref{tab:results} are overly conservative. Depending on its location, a false base station would be picked up by the reports of numerous mobile phones reporting to different serving cells and it is enough that \emph{one} of these serving cell models detects the malicious activity. This is illustrated in Table~\ref{tab:agg_results}, where each row contains the PCI of the false cell (instead of the serving cell as in Table~\ref{tab:results}). For each false cell, we categorize the false cell positions by the number of serving cells it is visible to, and for each such category the percentage of positions of the false cell detected as anomalous by at least one serving cell. 

\begin{table*}[h]
	\footnotesize
	\caption{Results per false cell id not including the static anomalies. Number of positions for the false cell (P) and ratio of detections (D) by at least one of the serving cell models listed for the detections of the ADF(\emph{COL}).}
\label{tab:agg_results}
\centering
  \resizebox{0.5\columnwidth}{!}{%
\begin{tabular}{|c|c|c|c|c|c|c|}
\hline
\hline

\multirow{2}{*}{False cell} & \multicolumn{2}{|c|}{Visible 1} &  \multicolumn{2}{|c|}{Visible 2} &  \multicolumn{2}{|c|}{Visible $>2$} \\
\cline{2-7}
 & P & D & P & D & P & D \\ 
\hline
\hline
	1        &  380  & 85.3\%&  96   & 84.4\%&   9   & 100.0\% \\
	2        &  445  & 73.5\%&  72   & 80.6\%&   1   & 100.0\% \\
	3        &  410  & 92.0\%&  149  & 99.3\%&   8   & 100.0\% \\
	4        &  543  & 81.8\%&  103  & 88.3\%&  10   & 90.0\% \\
	5        &  605  & 78.2\%&  166  & 92.8\%&  51   & 88.2\% \\ 
	6        &  593  & 85.7\%&  83   & 90.4\%&  24   & 95.8\% \\
	7        &  448  & 87.5\%&  113  & 99.1\%&  21   & 85.7\% \\
	8        &  460  & 76.5\%&  155  & 98.1\%&  36   & 100.0\% \\
	9        &  528  & 87.1\%&  84   & 90.5\%&  22   & 90.9\% \\
	10       &  479  & 82.9\%&  70   & 100.0\%&  10   & 100.0\% \\
	11       &  379  & 73.6\%&  59   & 84.7\%&   8   & 87.5\% \\
	12       &  365  & 83.6\%&  63   & 98.4\%&  10   & 100.0\% \\
\hline
\hline
\end{tabular}%
}
\end{table*} 

\section{Conclusion and Future Research}
We show that applying ML using RSRP-based features for network-based detection of false base station is promising. In our simulations with 12 cells where each cell alternated as a moving false cell, the best ML model obtained between 75-95\% recall of false cell's positions. Furthermore, analysis of the obtained anomaly scores indicates that a true false base station would cause a cluster of anomalies as multiple phones pick up the false cell in the measurements sent to different serving cells. This could make it feasible to distinguish such anomalies from false positives, for instance caused by an unusual phone location. We intend to continue the research, and validate our models and features on real operator data.

\clearpage
\bibliographystyle{unsrtnat}
\bibliography{./parts/ref}

\end{document}